\RequirePackage{fix-cm}
\RequirePackage{fixltx2e}
\documentclass[12pt,twoside,english,aip, jcp,floatfix]{revtex4-1}

\usepackage[T1]{fontenc}
\usepackage[latin9]{inputenc}
\usepackage{geometry}
\usepackage{xcolor}
\usepackage{babel}
\usepackage{amsmath}
\usepackage{amssymb}
\usepackage{graphicx}
\usepackage[unicode=true,pdfusetitle,
 bookmarks=false,
 breaklinks=true,pdfborder={0 0 0},pdfborderstyle={},backref=false,colorlinks=true]
 {hyperref}
\hypersetup{
 citecolor = blue, linkcolor = blue,  urlcolor  = blue}

\makeatletter
\renewcommand{\textendash}{--}

\usepackage{orcidlink}

\makeatother

\begin{document}

\title{Confidence curves for UQ validation: probabilistic reference vs.
oracle }

\author{Pascal PERNOT \orcidlink{0000-0001-8586-6222}}

\affiliation{Institut de Chimie Physique, UMR8000 CNRS,~\\
Université Paris-Saclay, 91405 Orsay, France}
\email{pascal.pernot@cnrs.fr}

\selectlanguage{english}%
\begin{abstract}
\noindent Confidence curves are used in uncertainty validation to
assess how large uncertainties ($u_{E}$) are associated with large
errors ($E$). An \emph{oracle} curve is commonly used as reference
to estimate the quality of the tested datasets. The oracle is a perfect,
deterministic, error predictor, such as $|E|=\pm u_{E}$, which corresponds
to a very unlikely error distribution in a probabilistic framework
and is unable unable to inform us on the calibration of $u_{E}$.
I propose here to replace the oracle by a probabilistic reference
curve, deriving from the more realistic scenario where errors should
be random draws from a distribution with standard deviation $u_{E}$.
The probabilistic curve and its confidence interval enable a direct
test of the quality of a confidence curve. Paired with the probabilistic
reference, a confidence curve can be used to check the calibration
and tightness of predictive uncertainties.\\
~\\
\\
\textbf{\textcolor{teal}{>\textcompwordmark{}>\textcompwordmark{}>
WARNING: This is a work in progress. }}\\
\textbf{\textcolor{teal}{>\textcompwordmark{}>\textcompwordmark{}>
All comments and suggestions are welcomed. }}
\end{abstract}
\maketitle

\section{Introduction}

Uncertainty quantification (UQ) is becoming a major issue for chemical
machine learning (ML),\citep{Weymuth2022} notably for the prediction
of molecular and material properties.\citep{Janet2019,Musil2019,Scalia2020,Tran2020,Wang2021,Zhan2021,Imbalzano2021,Busk2022}
As a corollary, UQ validation plays an essential role to assess the
\emph{calibration} of various UQ methods, and validation methods should
be chosen according to the nature of UQ metrics.\citep{Scalia2020,Pernot2022a,Pernot2022b}
UQ metrics quantify properties of the prediction errors, and one can
distinguish those targeting the \emph{amplitude} of errors from those
targeting the \emph{dispersion} of errors, typically through their
variance or quantiles. 

Amplitude metrics have been used a lot in computer vision,\citep{Kybic2011,MacAodha2012,Wannenwetsch2017,Ilg2018,Postels2022_arXiv}
but appear also in the physico-chemical context\citep{Korolev2022}.
The validation of amplitude metrics is mostly based on \emph{ranking-based}
methods, either \emph{correlation coefficients} between the metric
and absolute errors,\citep{Tynes2021} or so-called \emph{confidence
curves}\citep{Scalia2020}.\footnote{Confidence curves are also called a\emph{ sparsification error curve}
in computer vision.\citep{Ilg2018,Scalia2020}} 

To assess the quality of confidence curves, a best case scenario curve,
the \emph{oracle}, is usually plotted as reference. The distance between
a confidence curve and the oracle is estimated by statistics such
as the Area Under the Confidence-Oracle error (AUCO).\citep{Scalia2020}
As the oracle implies a perfect ranking between the amplitude metric
and the absolute errors, it corresponds to a correlation coefficient
equal to 1. A property of these ranking-based statistics is that they
do not depend on the scale of the amplitude metric. As such, they
cannot be used to estimate the \emph{calibration} of the metric.

Dispersion metrics are based on a probabilistic model of errors as
defined in metrology,\citep{GUM} and their validation should account
for this specificity. Typical validation methods compare the variance
of errors to the mean-squared prediction uncertainty or compare the
coverage of error prediction intervals to their target probability.\citep{Scalia2020,Pernot2022a,Pernot2022b}
The main difference with the amplitude metrics is that there is not
a symmetric relation between errors and uncertainty: large errors
are expected to be associated to large uncertainties, but small errors
can be associated to either small or large uncertainties. 

In this context, the use of ranking-based validation methods has limited
interest. For instance, sets of probabilistically valid errors and
uncertainties reach very modest correlation coefficients, around 0.5,
and one should not expect them to reach much larger values.\citep{Pernot2022b}
Similarly, the use of the oracle as a reference for confidence curves
does not make much sense. Still, ranking-based methods benefit of
some popularity in the chemistry ML-UQ community.\citep{Scalia2020,Tynes2021,Zheng2022,Korolev2022}
For instance, \emph{Scalia et al.}\citep{Scalia2020} used the oracle
and the associated distance metrics (AUCO) to compare variance-based
prediction uncertainties, and Korolev \emph{et al.}\citep{Korolev2022}
used it to compare variance-based uncertainties and their new amplitude-based
$\Delta$-metric.

One solution to this conflict would be to avoid the use of confidence
curves in the variance-based UQ framework. However, there are two
features of confidence curves that make them stand out from the other
methods used in calibration validation: (1) they provide a unique
and easy validation method for uncertainty metrics in active learning\citep{Pernot2022b};
and (2) they do not depend on a binning scheme which might introduce
some design problems in other validation methods (e.g., reliability
diagrams)\citep{Scalia2020}. 

In this article, I focus on the variance-based UQ framework and I
propose to switch from the \emph{oracle} to a \emph{probabilistic}
reference, and show that the resulting method, when carefully designed,
can be used as a validation tool for the calibration and \emph{tightness}
of variance-based uncertainties. The next section (Sect.\,\ref{sec:Confidence-curve,-oracle})
introduces the concepts and numerical methods and presents the options
to design a reliable probabilistic reference. Sect.\,\ref{sec:Application}
presents the application of the confidence curves to a selection of
datasets from the physico-chemical ML-UQ literature. These results
are discussed in Sect.\,\ref{sec:Discussion}.

\section{Confidence curve, oracle and probabilistic reference\label{sec:Confidence-curve,-oracle}}

Let us consider a validation dataset composed of \emph{paired} errors
and uncertainties $V=\left\{ E,u_{E}\right\} =\left\{ E_{i},u_{E_{i}}\right\} _{i=1}^{M}$
to be tested for calibration. In a recent article,\citep{Pernot2022b}
I reviewed the main variance-based UQ validation methods. These are
built on a probabilistic model
\begin{equation}
E_{i}\sim D(0,u_{E_{i}})\label{eq:probmod}
\end{equation}
linking errors to uncertainties, where $D(\mu,\sigma)$ is a probability
density function with mean $\mu$ and standard deviation $\sigma$.
This model states that errors should be unbiased ($\mu=0$) and that
uncertainties describe their dispersion, according to the metrological
standard.\citep{GUM}

The calibration of $u_{E}$ is based on testing that it correctly
describes the dispersion of $E$.\citep{Pernot2022a,Pernot2022b}
One can for instance test that
\begin{equation}
\mathrm{Var}(E)\simeq<u_{E}^{2}>\label{eq:varEVal}
\end{equation}
where the average is taken over the dataset $V$, and which is valid
only if errors are unbiased. However, this formula does not take into
account the pairing between errors and uncertainties, and a better
test can be written as
\begin{equation}
\mathrm{Var}(Z=E/u_{E})\simeq1\label{eq:varZval}
\end{equation}

The satisfaction of these tests validates the \emph{average} calibration,
which is a minimum requisite, but does not guarantee the reliability
of individual uncertainties. To achieve this, one can split $V$ into
subsets by binning $u_{E}$ and, within each subset, test Eq.\,\ref{eq:varEVal},
leading to \emph{reliability} \emph{diagrams}\citep{Levi2020}, or
test Eq.\,\ref{eq:varZval}, leading to the \emph{Local Z-Variance
analysis} (LZV)\citep{Pernot2022a,Pernot2022b}. 

In the following, I will refer to average calibration as \emph{calibration}
and to small-scale calibration as \emph{tightness}.\citep{Pernot2022b}
An ideal UQ method, i.e., a UQ method which provides reliable individual
uncertainties, should satisfy both calibration \emph{and} tightness.\citep{Pernot2022b} 

\subsection{Confidence curve\label{subsec:Confidence-curve}}

A confidence curve (CC) is established by estimating an error statistic
$S$ on subsets of $V$ pruned from the points with uncertainties
larger than a threshold.\citep{Scalia2020} It is also called a\emph{
sparsification error curve} in computer vision.\citep{Ilg2018,Scalia2020}
Technically, it is a ranking-based method, as (1) it is insensitive
to the scale of the uncertainties, and (2) the relative ordering of
the errors and uncertainties plays a determinant role. 

If one defines the threshold $u_{k}$ as the largest uncertainty after
removing the $k$\,\% largest uncertainties from $u_{E}$ ($k\in\{0,1,\ldots,99\}$),
a confidence statistic is defined by
\begin{equation}
c_{S}(k;E,u_{E})=S\left(E\,|\,u_{E}<u_{k}\right)\label{eq:non-normcc}
\end{equation}
and its normalized version by
\begin{equation}
\tilde{c}_{S}(k;E,u_{E})=c_{S}(k;E,u_{E})/S(E)\label{eq:conf}
\end{equation}
where $S$ is an error statistic \textendash{} typically the Mean
Absolute Error (MAE) or Root Mean Squared Error (RMSE) \textendash{}
and $S\left(E\,|\,u_{E}<u_{k}\right)$ denotes that only those errors
$E_{i}$ paired with uncertainties $u_{E_{i}}$ smaller than $u_{k}$
are selected to compute $S$. A confidence curve is obtained by plotting
$c_{S}$ or $\tilde{c}_{S}$ against $k$. Both normalized and non-normalized
CCs are used in the literature. 

A continuously decreasing CC reveals a desirable association between
the larger errors and the larger uncertainties, an essential feature
for active learning or to detect unreliable predictions. 

\subsection{Reference curves\label{subsec:Reference-curves}}

As already mentioned, only the order of $u_{E}$ values is used in
a CC, and any change of scale of $u_{E}$ leaves $c_{s}$ unchanged.
Without a proper reference, $c_{S}$ cannot inform us on calibration
or tightness. One should note also that the normalized version, $\tilde{c}_{S}$,
is also insensitive to scale changes of $E$, which excludes it for
any calibration validation application, whichever the chosen reference,
unless it is complemented by another calibration test. 

\subsubsection{Oracle \label{subsec:Oracle}}

An oracle curve can be generated from the validation dataset $V$
by reordering $u_{E}$ to match the order of absolute errors. This
can be expressed as
\begin{equation}
O(k;E)=c_{s}(k;E,|E|)
\end{equation}
with a similar expression for the normalized version. 

It is evident from the above equation that the oracle is independent
of $u_{E}$ and therefore useless for calibration testing. Recast
in the probabilistic framework introduced above, the oracle would
correspond to a very implausible error distribution $D$, such that
$E_{i}=\pm u_{E_{i}}$. 

\subsubsection{Probabilistic reference\label{subsec:Probabilistic-reference}}

Using Eq.\,\ref{eq:probmod}, a probabilistic reference curve $P$
can be generated by sampling pseudo-errors $\widetilde{E}_{i}$ for
each uncertainty $u_{E_{i}}$ and calculating a CC for $\left\{ \tilde{E},u_{E}\right\} $,
i.e.,
\begin{equation}
P(k;u_{E})=\left\langle c_{S}(k;\tilde{E},u_{E})\right\rangle _{\tilde{E}}
\end{equation}
where a Monte Carlo average is taken over samples of
\begin{equation}
\widetilde{E}_{i}\sim D(0,u_{E_{i}})
\end{equation}
The sampling is repeated a number of times sufficient to have a stable
mean and confidence band (at the 95\,\% level). 

In contrast to the oracle, which depends exclusively on the errors,
the probabilistic reference depends on $u_{E}$ and a choice of distribution
$D$. It does not depend on the actual errors $E$. Comparison of
the data CC to $P$ enables to test if $E$ and $u_{E}$ are correctly
linked by the probabilistic model, Eq.\,\ref{eq:probmod}, i.e. to
test if the uncertainties quantify correctly the dispersion of the
errors.

\paragraph{Examples.}

\begin{figure}[t]
\noindent \begin{centering}
\includegraphics[height=8cm]{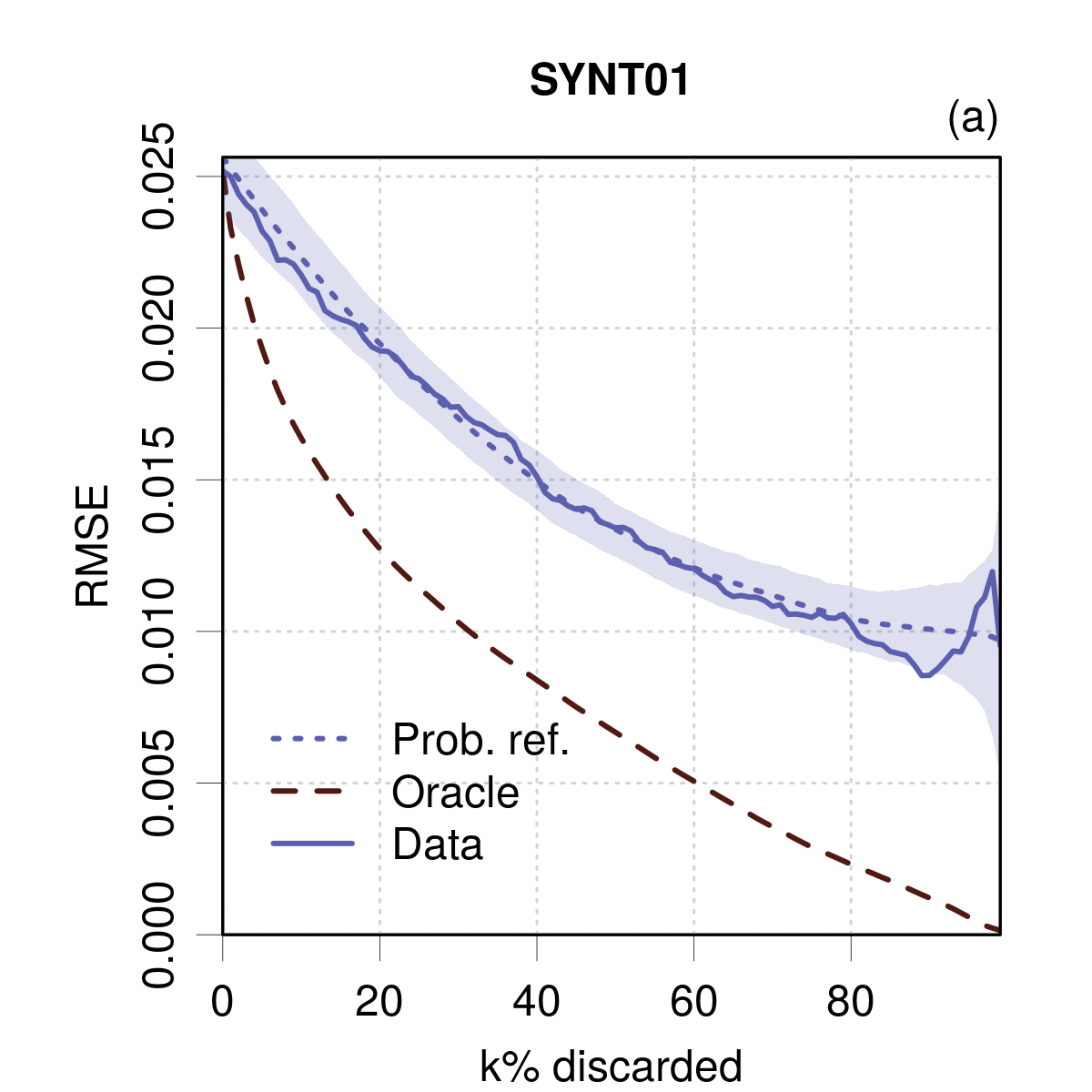}\includegraphics[height=8cm]{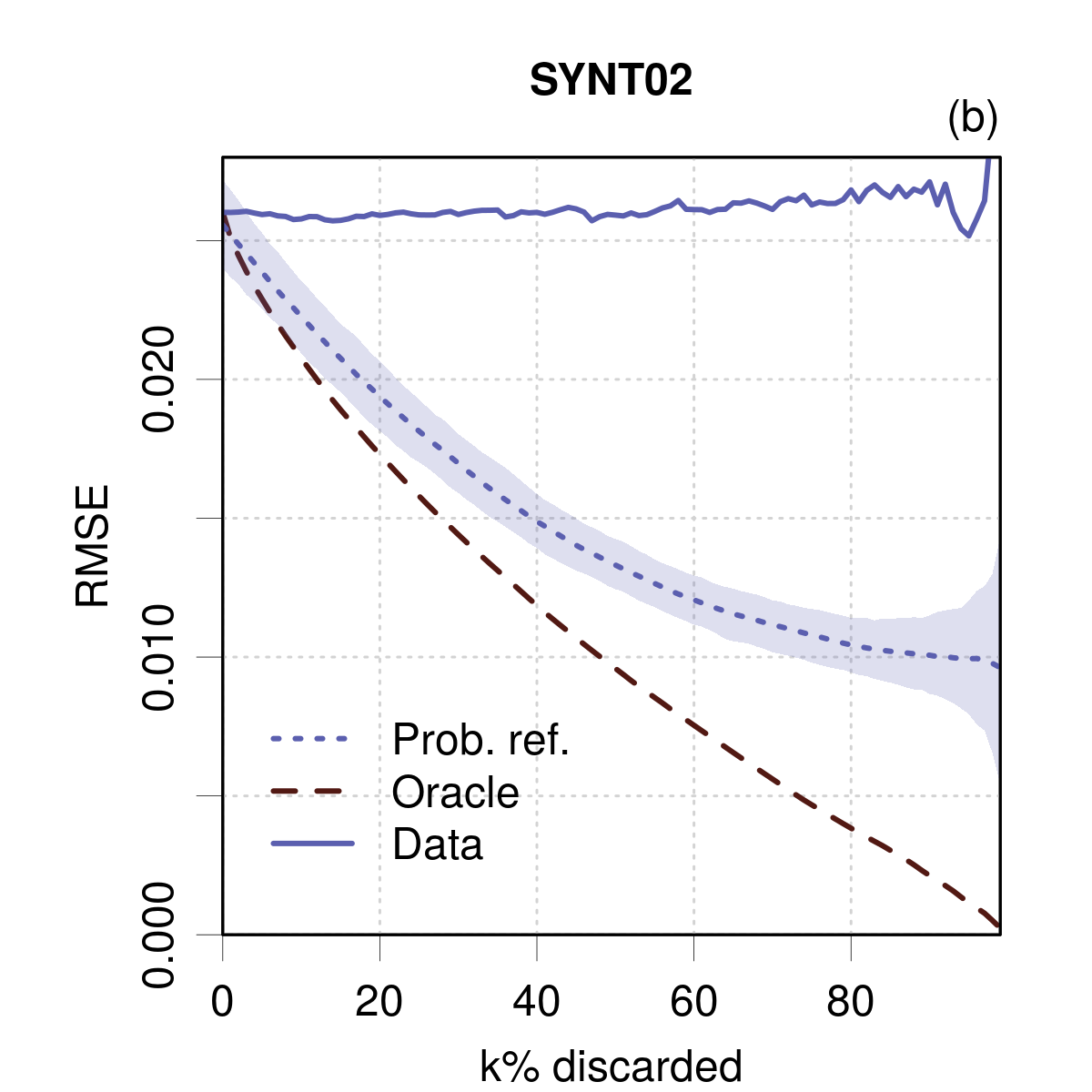}
\par\end{centering}
\caption{\label{fig:Examples-of-confidence}Examples of RMSE-based confidence
curves for two synthetic datasets: (a) SYNT01 is a dataset where the
errors are conform to the probabilistic model Eq.\,\ref{eq:probmod};
(b) SYNT02 is a dataset of errors sampled from a probabilistic model
with constant uncertainty. }
\end{figure}

A comparison of both reference curves for synthetic datasets is shown
in Fig.\,\ref{fig:Examples-of-confidence}. The statistic $S$ is
the RMSE. The probabilistic reference and its 95\,\% confidence band
have been obtained from samples of 500 realizations.

SYNT01 is a dataset generated using Eq.\,\ref{eq:probmod} and a
normal distribution ($M=1000$). Its CC is conform to the probabilistic
reference, confirming the validity of the uncertainties to describe
the dispersion of the errors. It is impossible to get such information
by using the oracle curve as a reference. The oracle curves decrease
to zero because the smallest errors (even though they might originate
randomly from large uncertainties) are associated by the oracle to
the smallest uncertainties. Note that the confidence band of the probabilistic
reference widens notably for large $k$ values, as the size of the
remaining set in this area might get small. In this case, the last
bin ($k=99$) contains only $M/100=10$ pairs. 

The SYNT02 dataset {[}Fig.\,\ref{fig:Examples-of-confidence}(b){]}
uses the same uncertainties as SYNT01 (it has therefore the same probabilistic
reference), but the errors are now generated from a normal distribution
with a \emph{constant} standard deviation. The absence of link between
errors and uncertainties is clearly revealed by the non-decreasing
confidence curve. 

\paragraph{Effect of the generative distribution and error statistic.}

The probabilistic reference $P$ depends on the choice of a generative
distribution $D$ and of an error statistic $S$ (which should be
the same as for the CC itself). I show here how these two factors
might interfere and how to minimize the dependence of $P$ on $D$.

A default choice for $D$ would be the normal distribution which is
often assumed in ML-UQ calibration studies. However, the error budget
is very often dominated by \emph{model errors}, which have no reason
to be normally distributed.\citep{Pernot2022a} In consequence, I
consider four other candidate distributions covering a wide range
of kurtosis values: Uniform ($\kappa=1.8$), Exponential power distribution
with $p=4$ (Normp4; $\kappa=2.2$), Laplace ($\kappa=6.0$), and
Student's-\emph{t} with four degrees of freedom (T4; $\kappa=\infty$).\citep{Pernot2022b}

\begin{figure}[t]
\noindent \begin{centering}
\includegraphics[height=8cm]{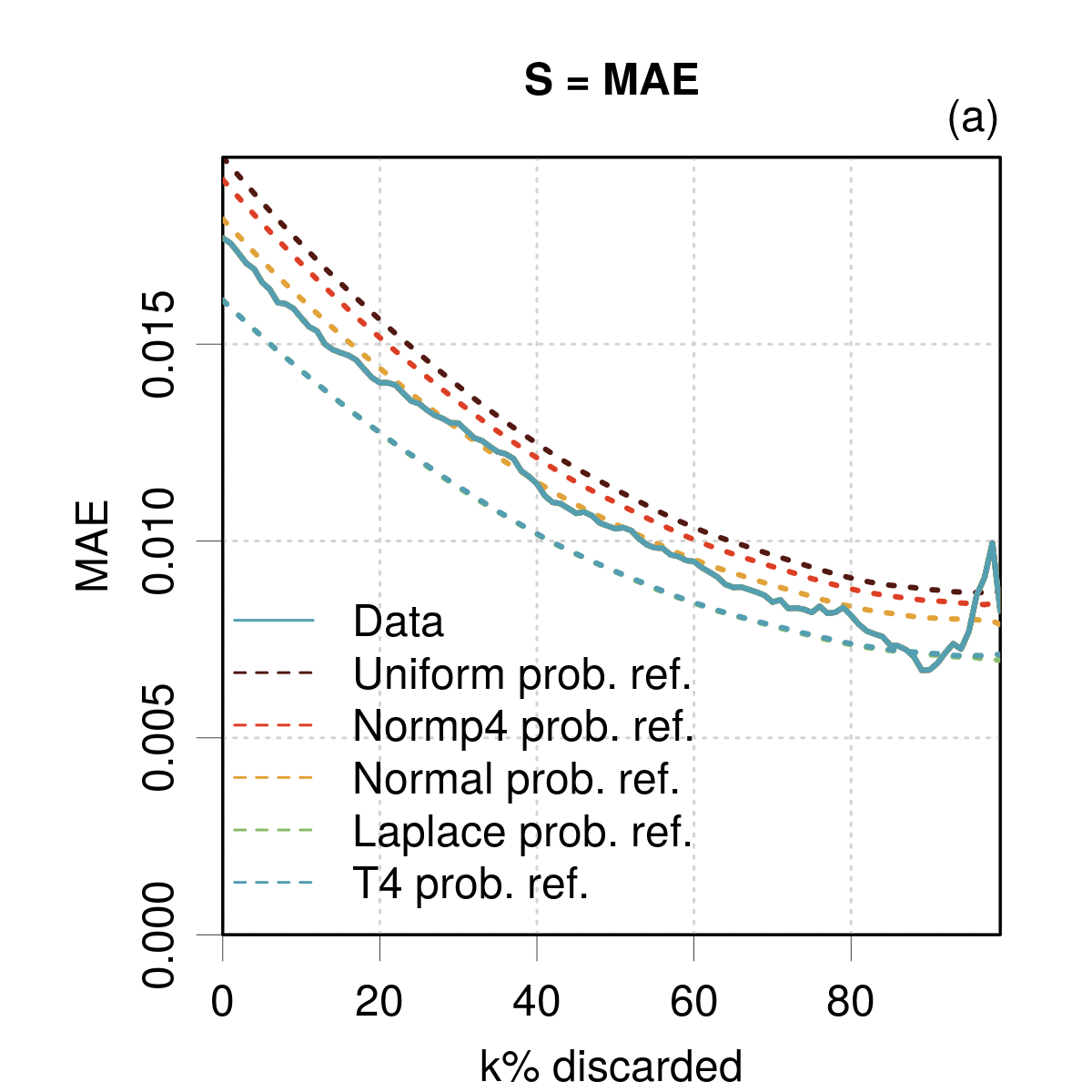}\includegraphics[height=8cm]{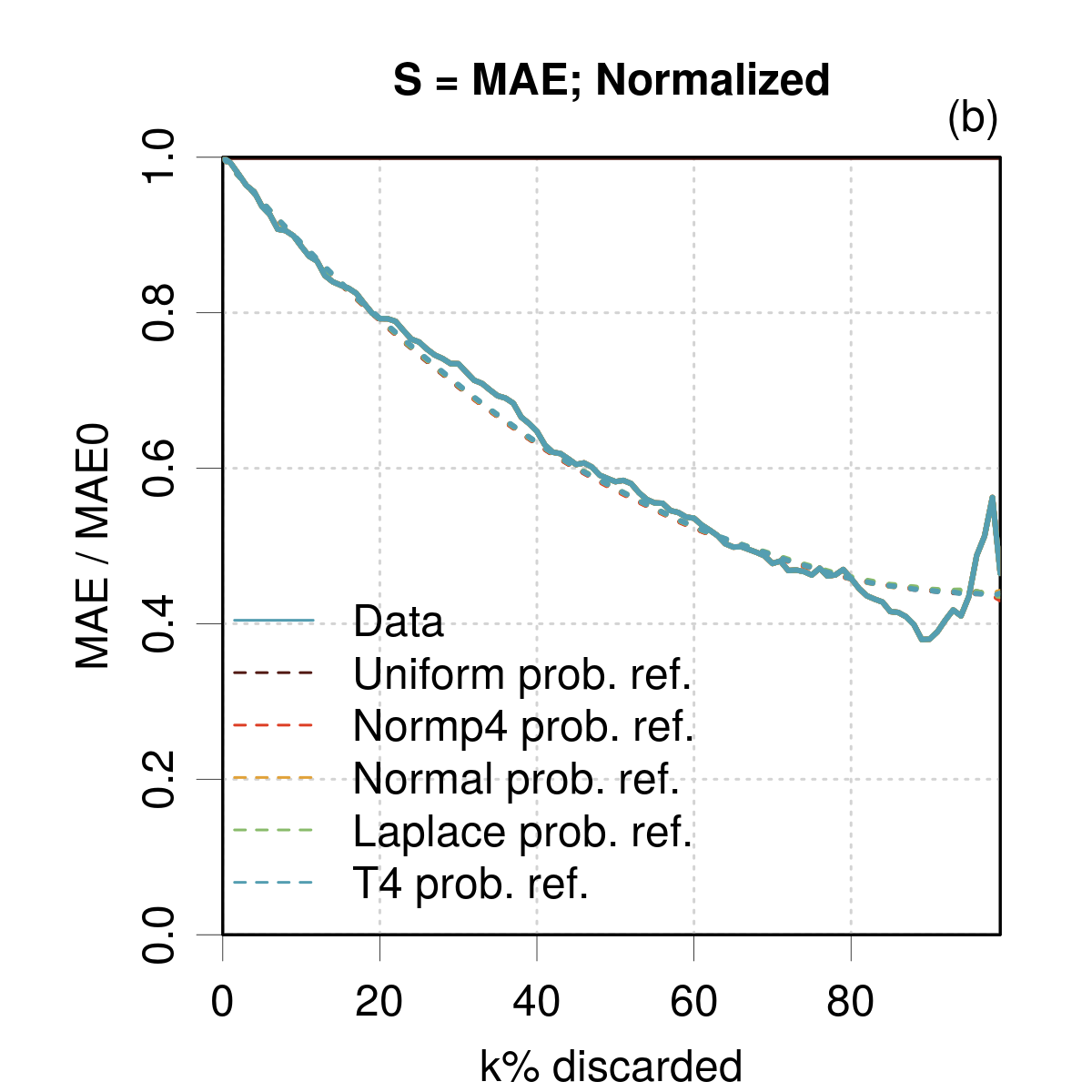}
\par\end{centering}
\noindent \begin{centering}
\includegraphics[height=8cm]{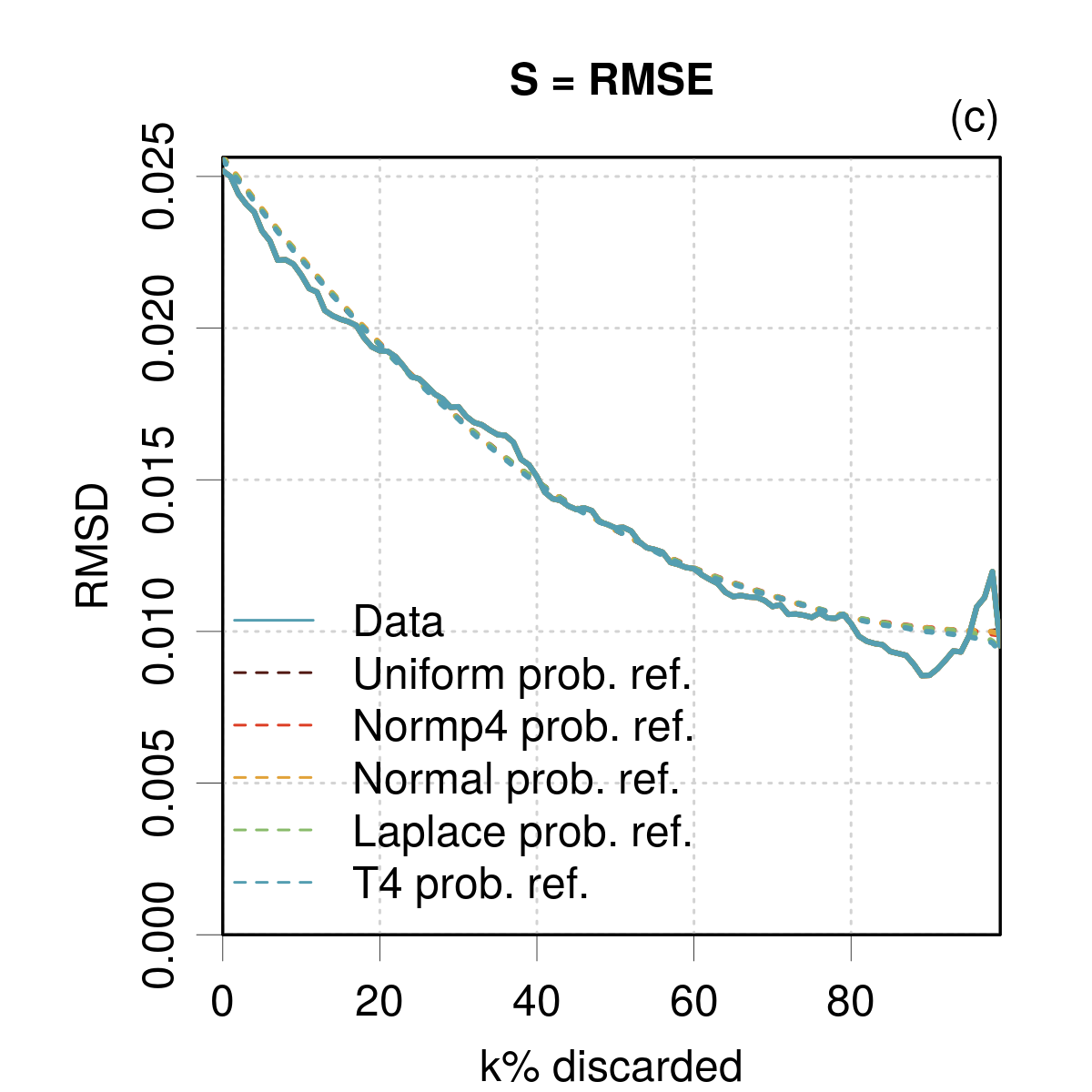}\includegraphics[height=8cm]{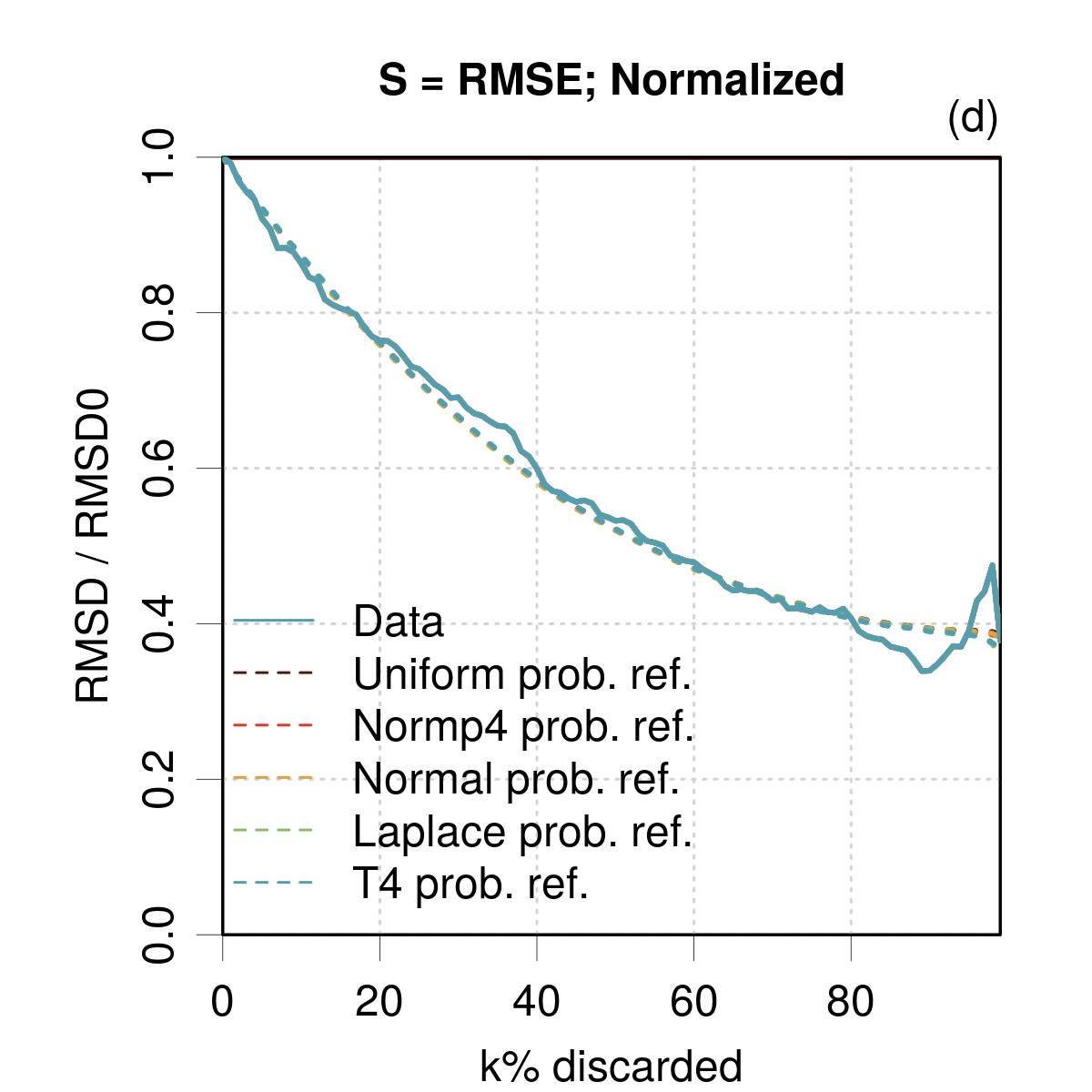}
\par\end{centering}
\caption{\label{fig:impact-dist}Impact of the probabilistic model distribution
on the reference curve for non-normalized (a,c) and normalized (b,d)
confidence curves and two statistics, MAE (a,b) and RMSD (c,d)}
\end{figure}

The corresponding probabilistic reference curves for the SYNT01 dataset
are reported in Fig.\,\ref{fig:impact-dist}, for $c_{S}$ and $\tilde{c}_{S}$
and two $S$ statistics (MAE and RMSE). In all cases, except for the
non-normalized CC using the MAE, the reference curves are overlapping.
As the pseudo-errors are generated by a distribution with prescribed
standard deviation, the RMSE-based reference curves are not sensitive
to the choice of $D$. This is not the case for the MAE-based reference
curves (note that the Laplace and T4 curves are overlapping). However,
this difference is fully compensated by normalization. It appears
thus that for non-normalized CCs, it is better to use the RMSE and
benefit from the insensitivity of the corresponding probabilistic
reference on $D$. 

\begin{figure}[t]
\noindent \begin{centering}
\includegraphics[height=8cm]{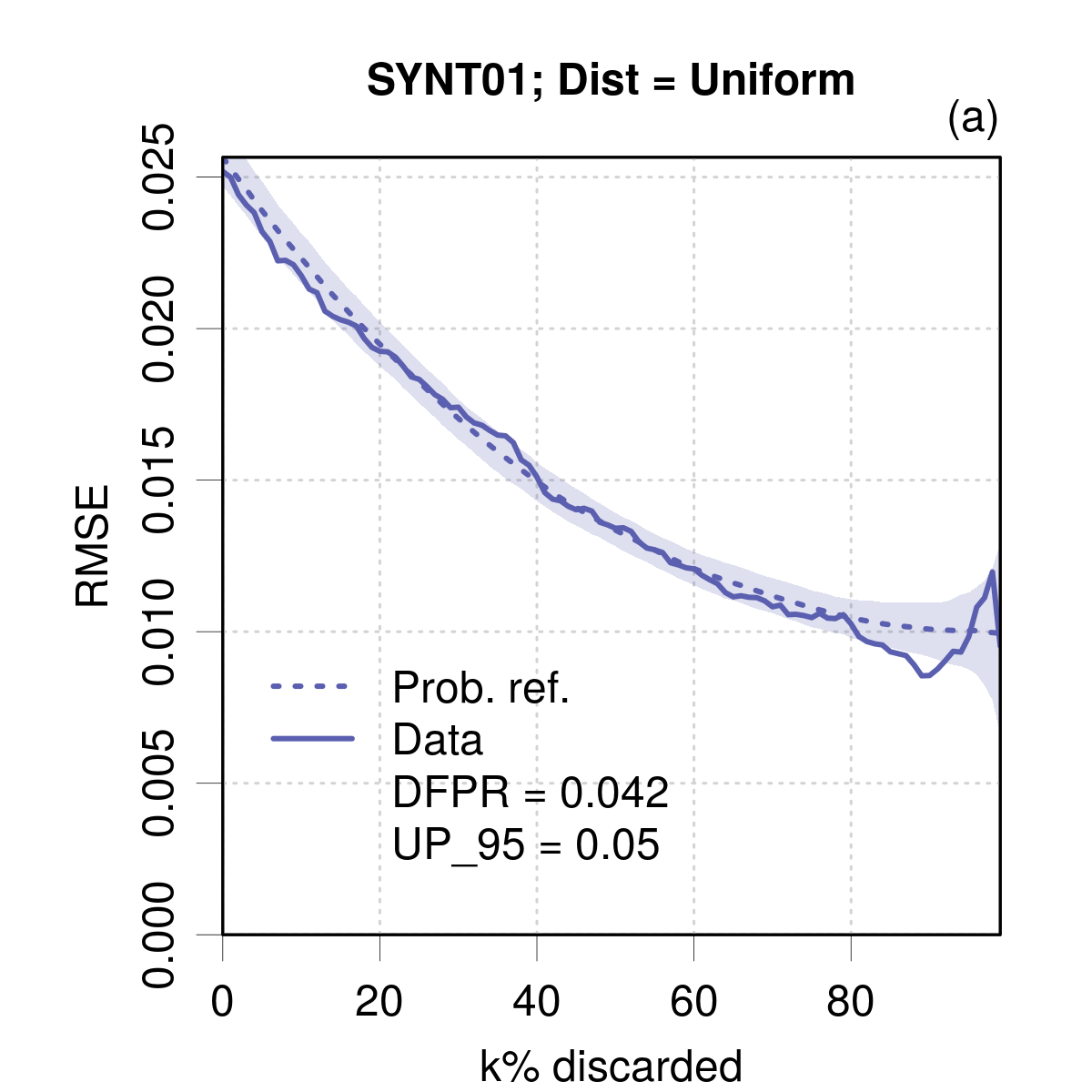}\includegraphics[height=8cm]{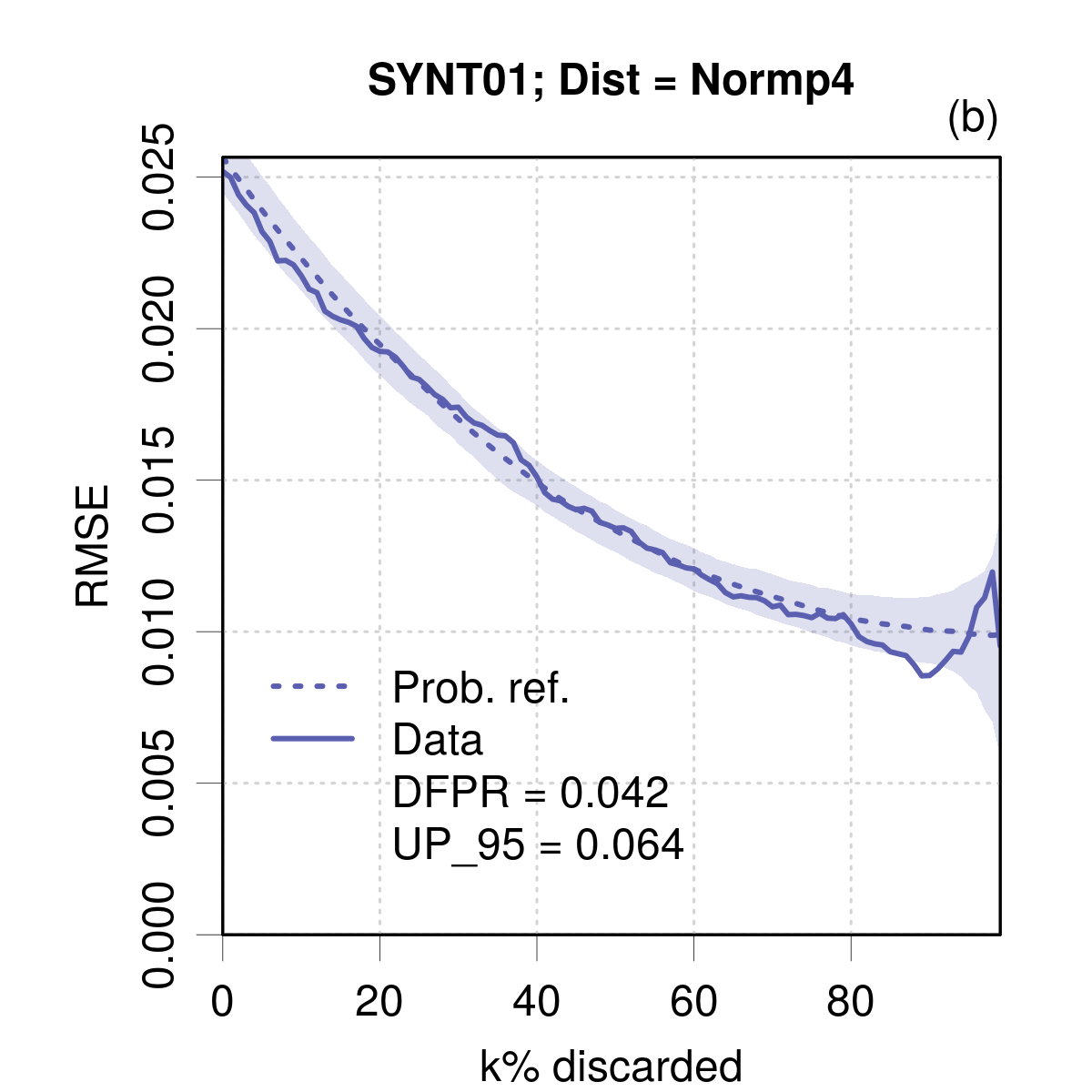}
\par\end{centering}
\noindent \begin{centering}
\includegraphics[height=8cm]{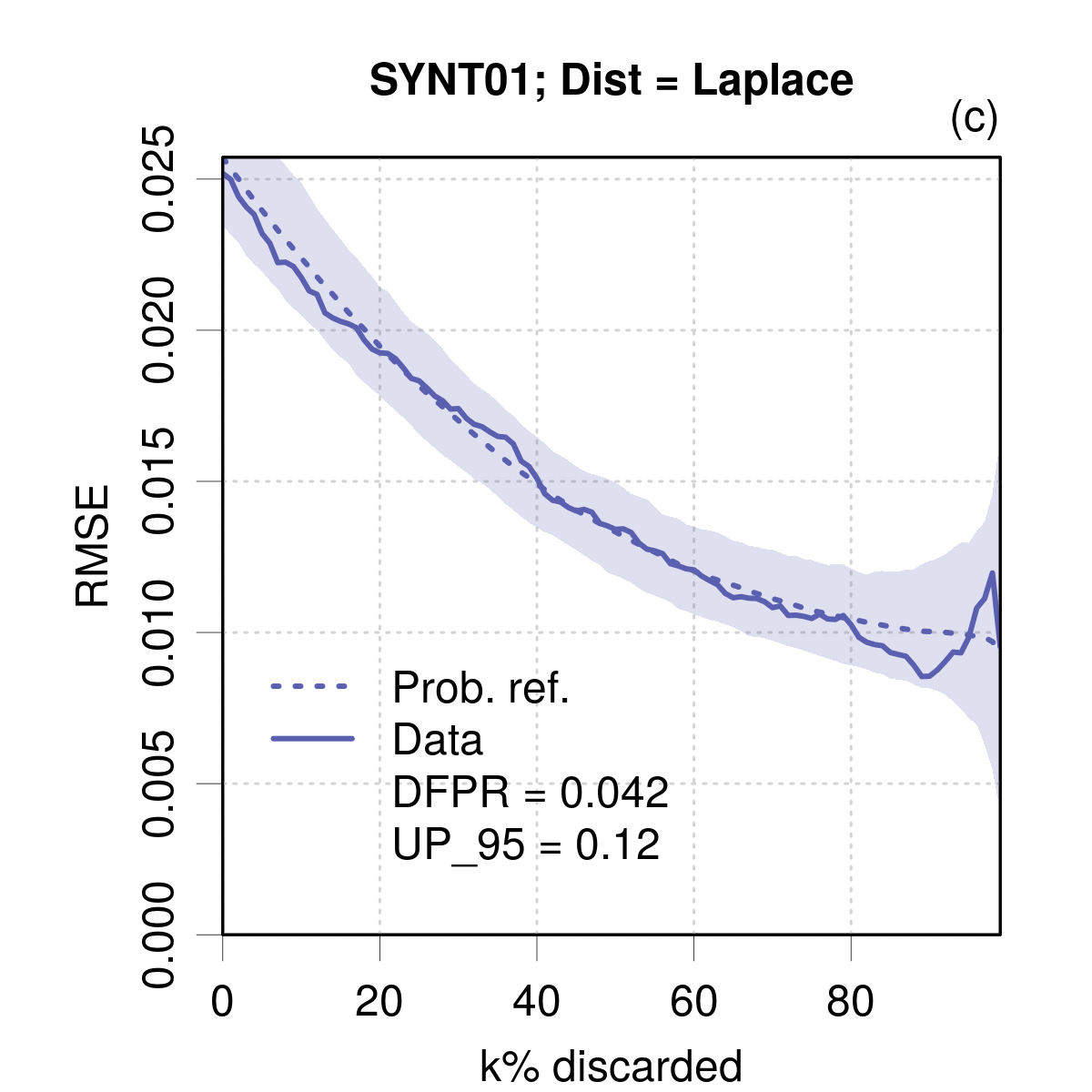}\includegraphics[height=8cm]{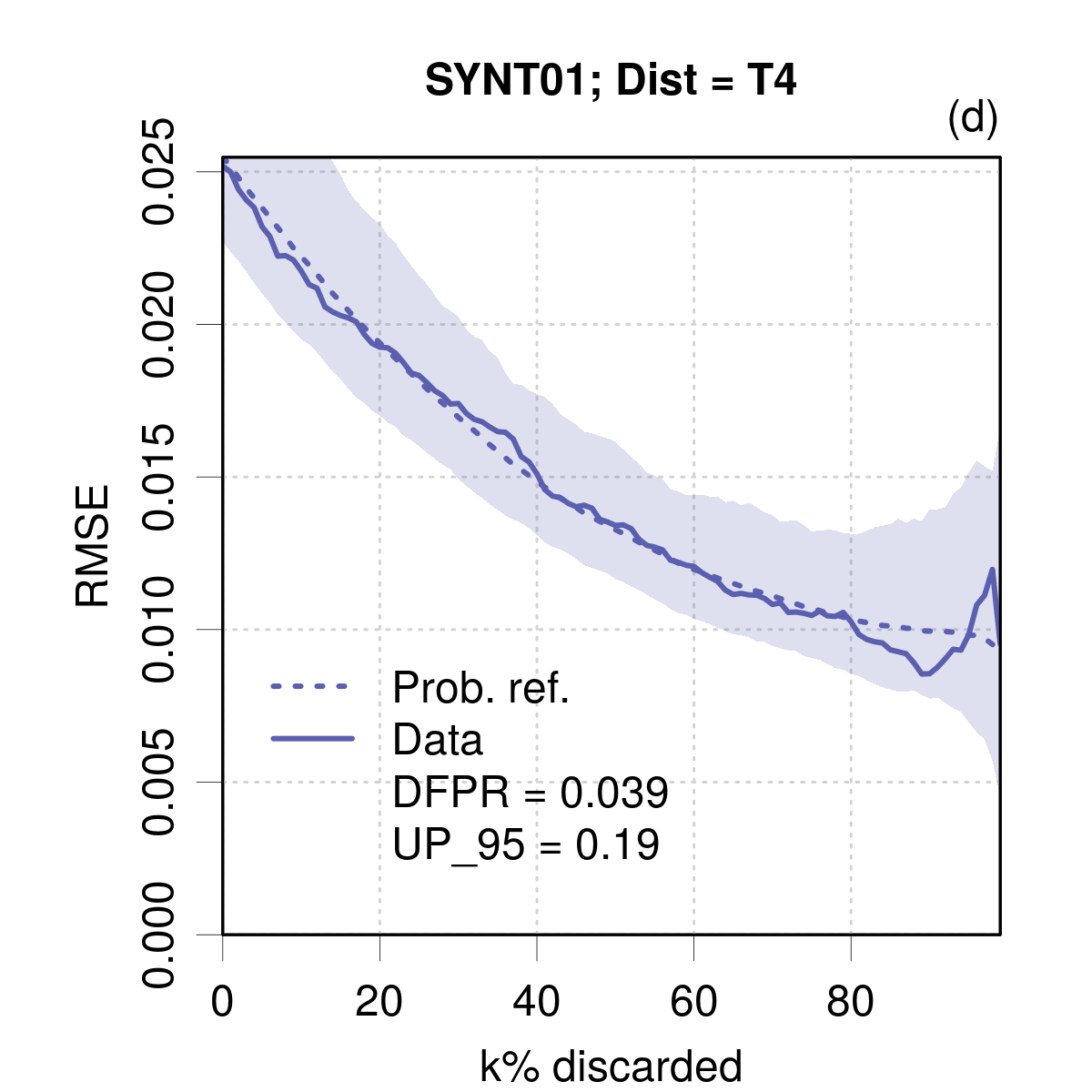}
\par\end{centering}
\caption{\label{fig:impact-dist-1}Impact of the probabilistic model distribution
on the confidence band of the reference curve for non-normalized confidence
curves. For the Normal case, see Fig.\,\ref{fig:Examples-of-confidence}(a).}
\end{figure}

Let us now consider the confidence bands of the probabilistic reference
due to the choice of $D$, given the RMSE statistic {[}Fig.\,\ref{fig:impact-dist-1}
and Fig.\,\ref{fig:Examples-of-confidence}(a){]}. One can see a
notable effect of $D$ on the width of the confidence band: distributions
with higher kurtosis values lead to wider confidence bands. This has
to be kept in mind, but it can be considered as a secondary order
effect that can be discussed when a CC lies close to the reference.
In absence of specific information about the generative distribution
$D$, the normal provides thus a balanced default choice. 

\subsection{Distance from the probabilistic reference\label{subsec:Distance-from-the}}

In the same spirit as the AUCO,\citep{Scalia2020} the distance of
the confidence curve from its probabilistic reference can be used
as a calibration/tightness metric
\begin{equation}
\mathrm{DFPR}=\sum_{i=0}^{99}|c_{S}(i;E,u_{E})-P(i;u_{E})|
\end{equation}
For the validation of the DFPR, a threshold UP95 is estimated as the
95th quantile of the distribution of DFPR values generated by random
sampling of the probabilistic model. If $\mathrm{DFPR<UP95}$ one
can conclude that the uncertainties and errors are in agreement with
Eq.\,\ref{eq:probmod}. Note that the value of UP95 depends on the
choice of generative distribution $D$ {[}Fig.\,\ref{fig:impact-dist-1}{]},
as is the case for the confidence bands of the probabilistic reference,
and its use comes with the same caveats as for the confidence bands. 

In order to avoid interpretation confusions, I do not define this
metric for normalized confidence curves. 

\section{Applications\label{sec:Application}}

\noindent In a recent article, Palmer \emph{et al.}\citep{Palmer2022}
proposed a \emph{calibrated bootstrap} method to improve UQ resulting
from various ML methods. They provide as supplementary information
20 validation datasets with errors and uncertainties for uncalibrated
and calibrated results. I do not consider here the synthetic ones,
but focus on those based on physical data, i.e. the Diffusion\citep{Lu2019}
($M=2040$) and Perovskite\citep{Li2018} ($M=3836$) datasets. I
want to emphasize that it is not my intent here to evaluate the \emph{calibrated
bootstrap} method, but only to show how the confidence curves can
be used in replacement or as complement to a LZV analysis. 

\begin{figure}[t]
\noindent \begin{centering}
\includegraphics[height=8cm]{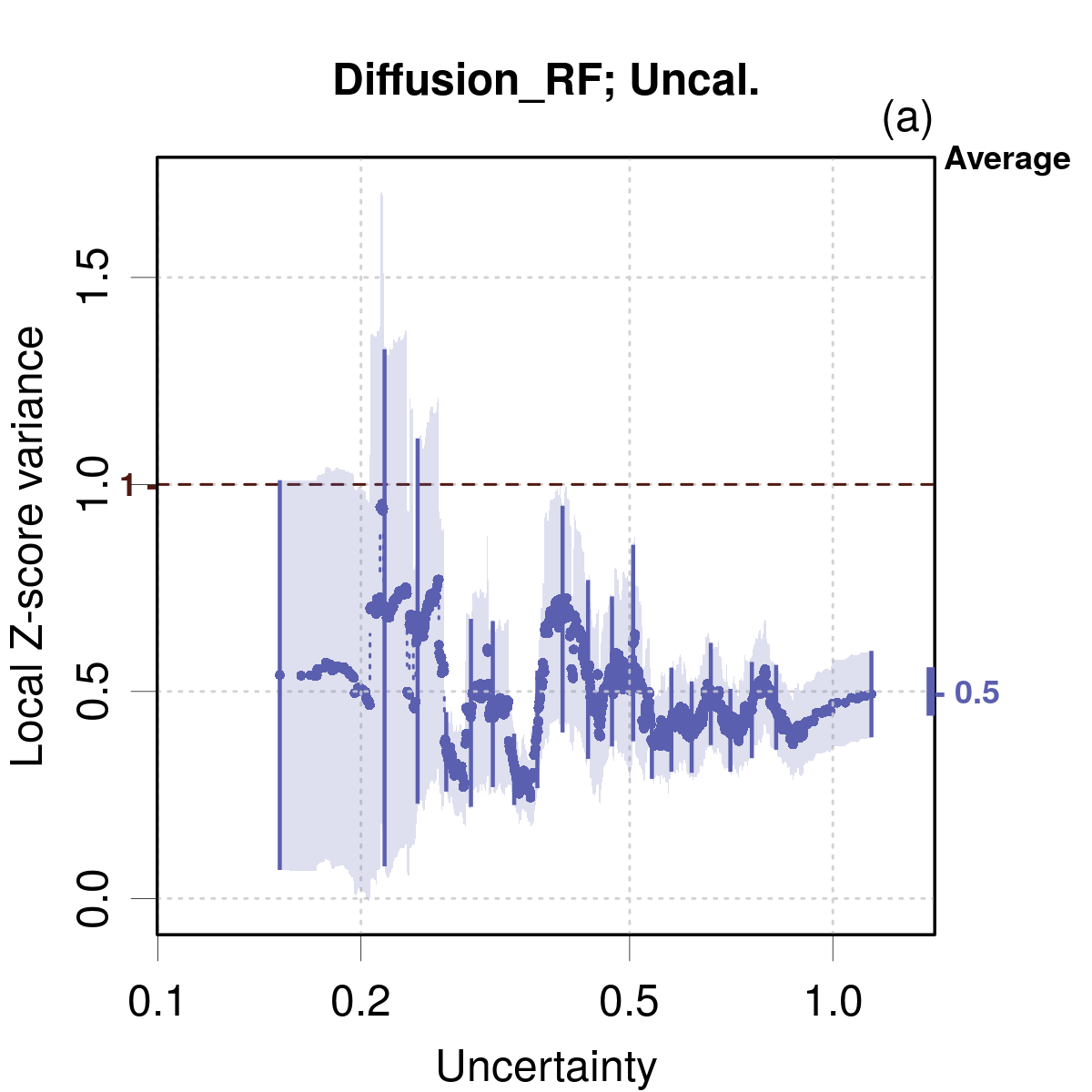}\includegraphics[height=8cm]{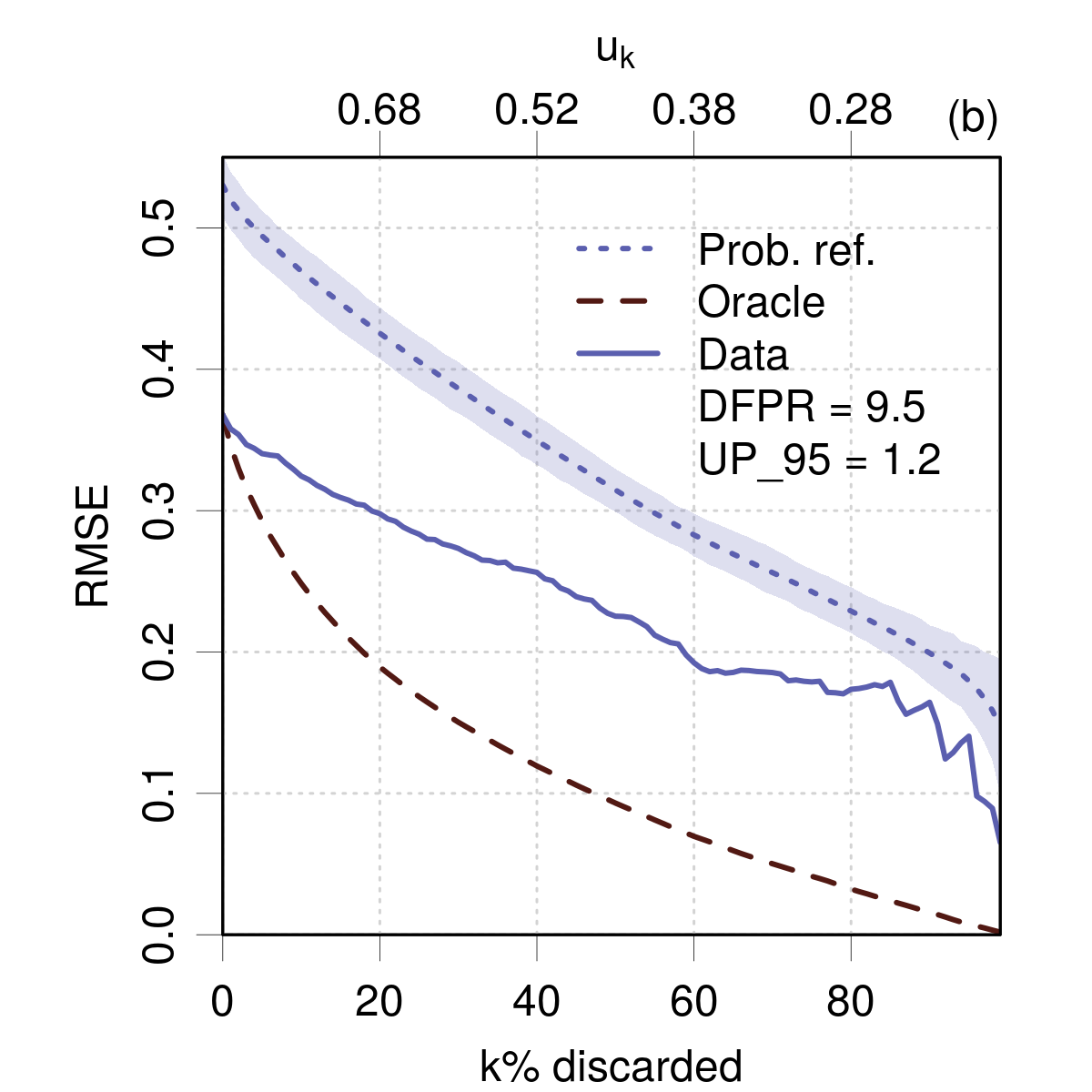}
\par\end{centering}
\noindent \begin{centering}
\includegraphics[height=8cm]{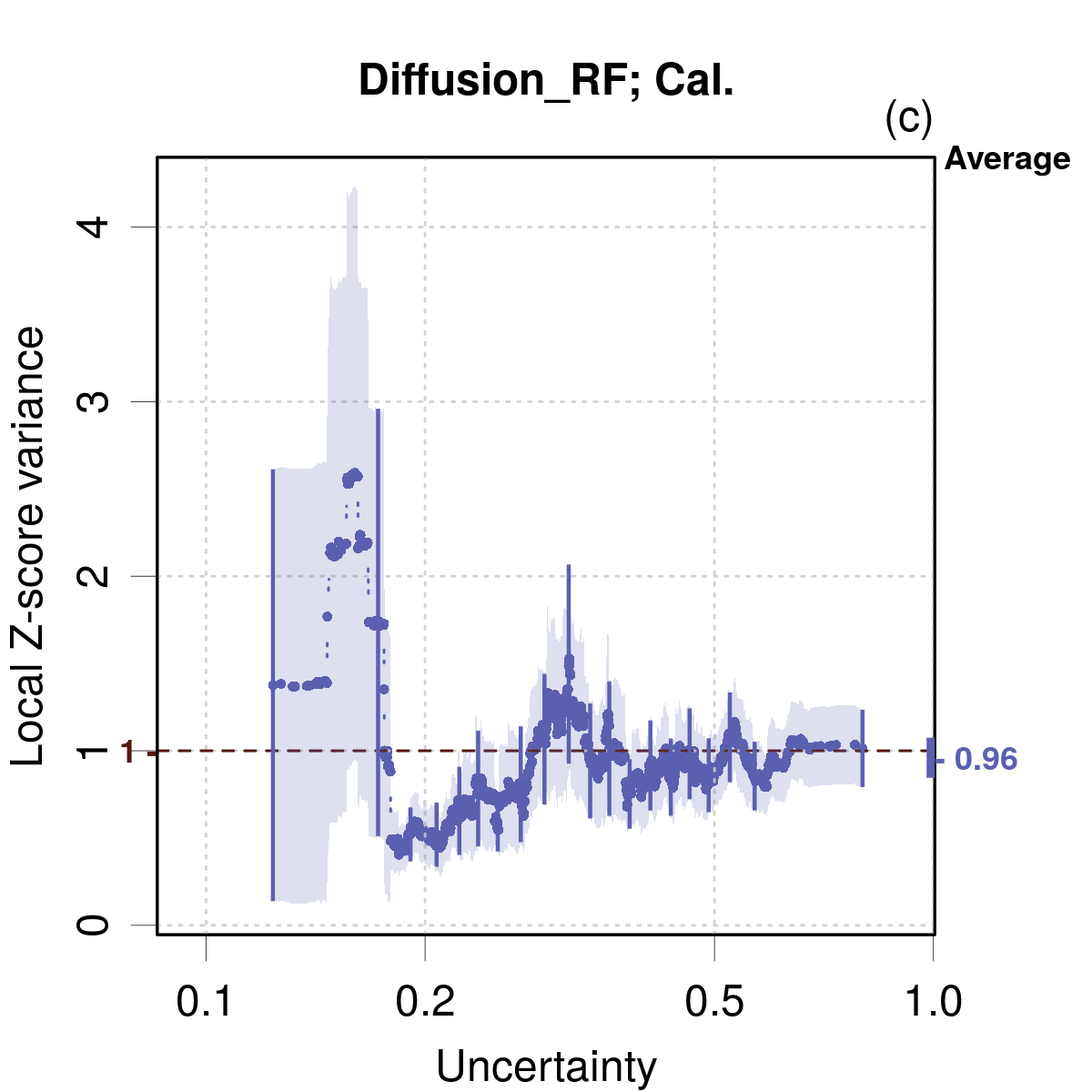}\includegraphics[height=8cm]{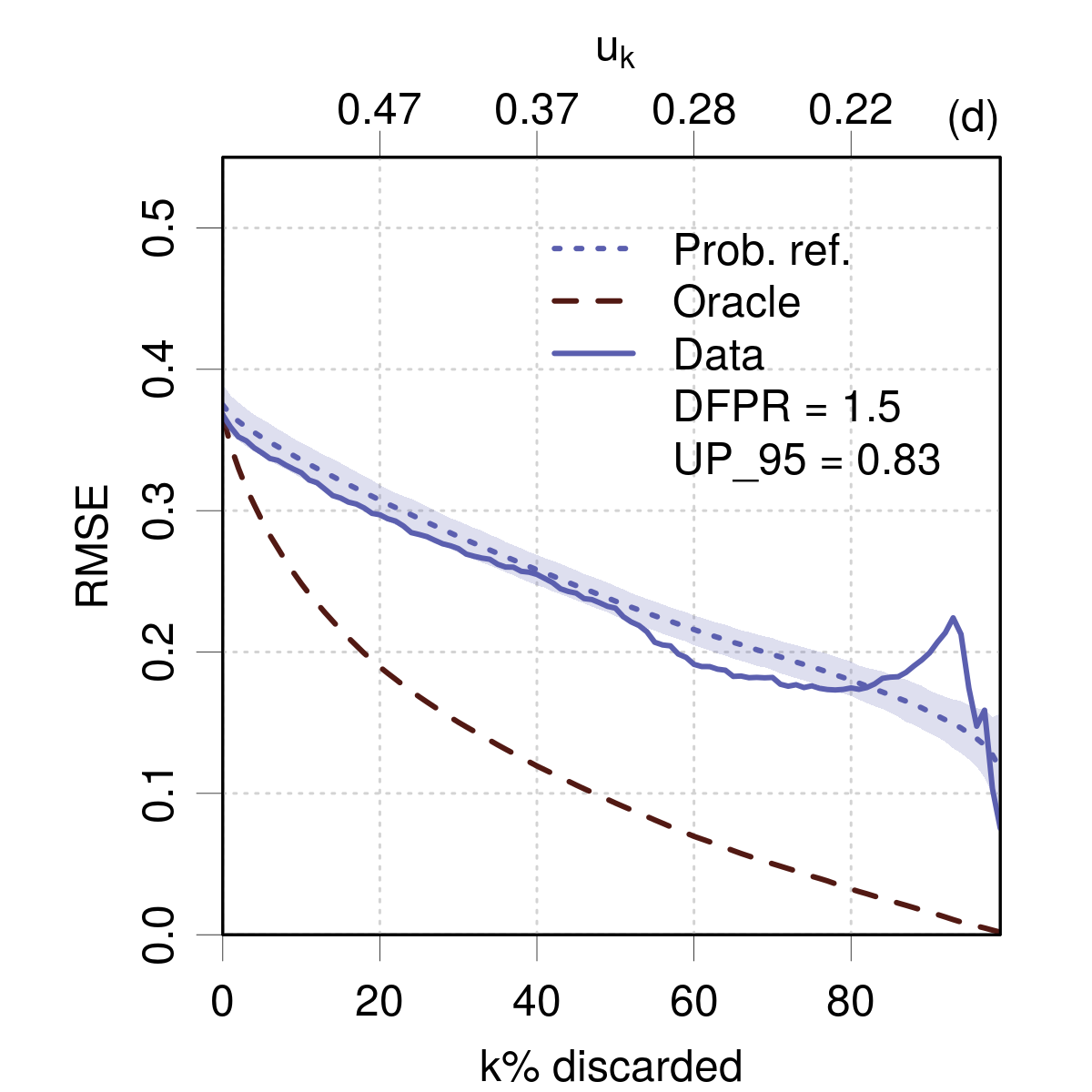}
\par\end{centering}
\caption{\label{fig:PAL1-0}LZV analysis (a,c) and confidence curves (b,d)
for uncalibrated (a,b) and calibrated (c,d) uncertainties from the
Diffusion\_RF dataset of Palmer \emph{et al.}\citep{Palmer2022}.
Threshold uncertainties $u_{k}$ for $k=$ 20, 40, 60 and 80 have
been added to the confidence plots to facilitate the comparison with
LZV plots.}
\end{figure}

In Fig\,\ref{fig:PAL1-0}, I present the results of the LZV analysis
and confidence curves for the Random Forest application to the Diffusion
dataset (denoted Diffusion\_RF). For the uncalibrated results {[}Figs\,\ref{fig:PAL1-0}(a,b){]},
one sees that average calibration is off ($\mathrm{Var}(Z)=0.5$),
indicating an average overestimation of the uncertainties by a factor
around $\sqrt{1/0.5}=\sqrt{2}$. The CC lies \emph{below} the probabilistic
reference, indicating overestimation of the uncertainties, in agreement
with the LZV analysis. The DFPR statistic ($9.5$) is largely above
the UP95 limit ($1.1$). 

The calibrated bootstrap results {[}Figs\,\ref{fig:PAL1-0}(c,d){]}
show that average calibration has been reached $\mathrm{Var}(Z)=0.96$,
but at the cost of large local deviations for the smaller uncertainties.
Both graphs indicate a good tightness for uncertainties above $u_{E}\simeq0.3$
(about 50\,\% of the dataset) and present mirrored deviations below
this value. The DFPR statistics ($1.5$) is still above UP95 ($0.91$),
but much less than before calibration. 

By comparing the confidence curves for uncalibrated and calibrated
data, one sees that the confidence curve of the calibrated data does
not get closer to the oracle. Moreover, the calibration curves are
very similar, except for the smaller uncertainties, indicating that
the calibration process preserves the ranks of the 60-70\,\% larger
uncertainties. By contrast, the probabilistic reference is sensitive
to calibration and lies much closer to the confidence curve of calibrated
data.

Interestingly, the LZV curve for the uncalibrated data oscillates
around a constant value, suggesting that a simple scaling of the uncertainties
might provide a notable improvement. In this case, a simple \emph{a
posteriori} calibration by division of the uncalibrated uncertainties
by a factor $\sqrt{2}$ provides calibrated ($\mathrm{Var}(Z)=1$)
and fairly tight uncertainties (not shown), with a DFPR value ($0.77$)
smaller than UQ95 ($0.83$). However, this \emph{ad hoc} scaling would
not be successful for the other datasets presented below. 

The Diffusion\_RF example is representative of the efficiency the
calibrated bootstrap.\citep{Palmer2022} In the following, I focus
on the analysis of calibrated data.

\begin{figure}[t]
\noindent \begin{centering}
\includegraphics[height=8cm]{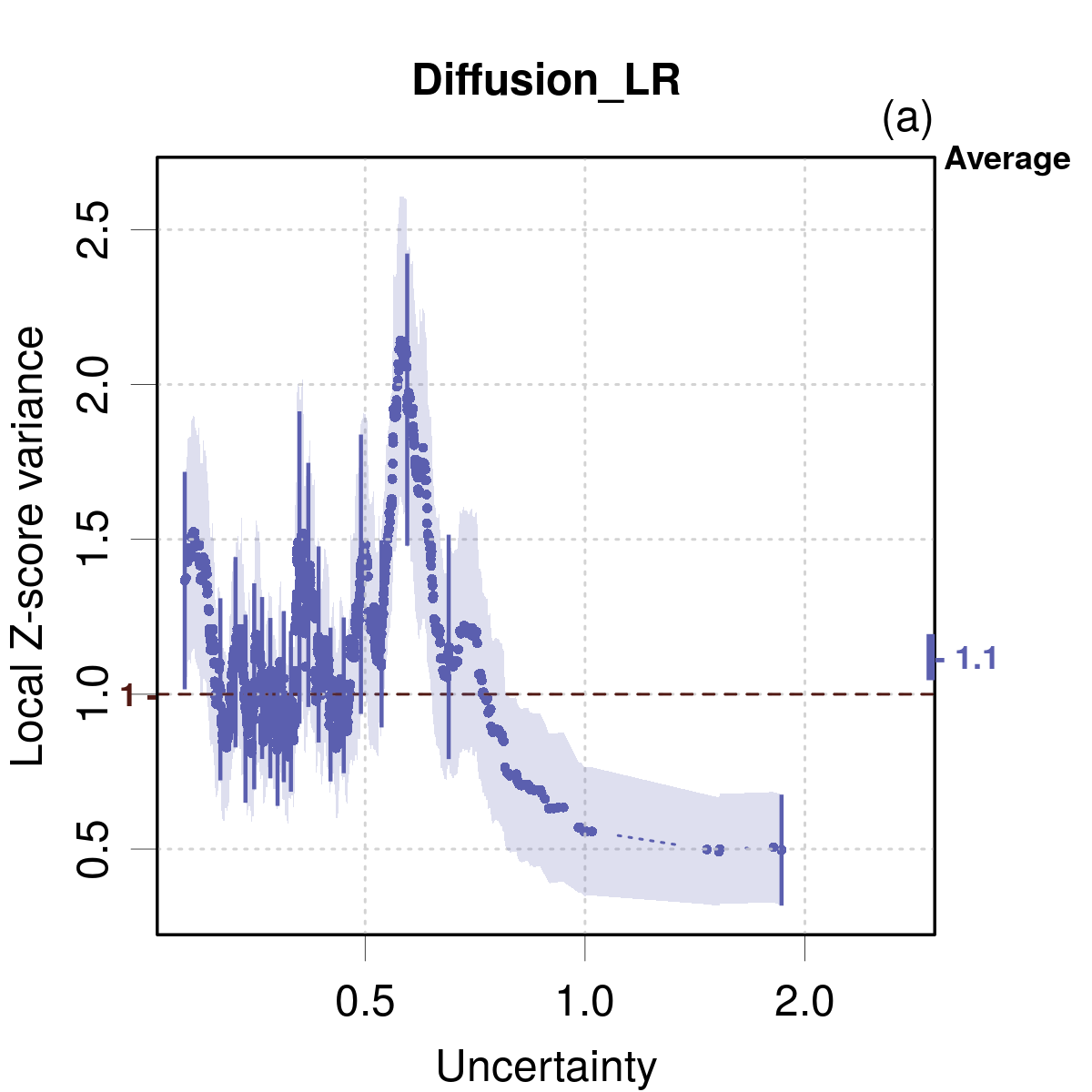}\includegraphics[height=8cm]{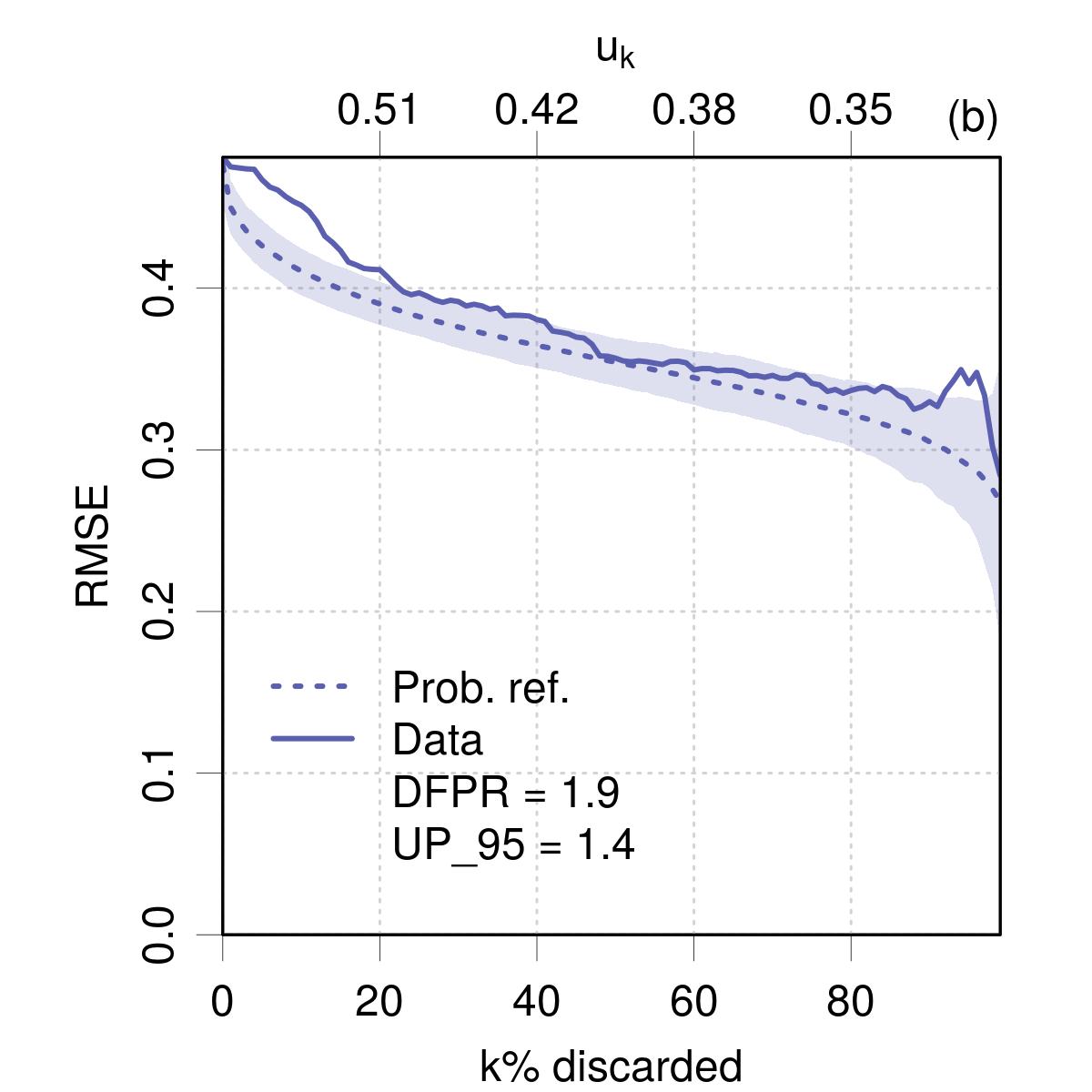}
\par\end{centering}
\noindent \begin{centering}
\includegraphics[height=8cm]{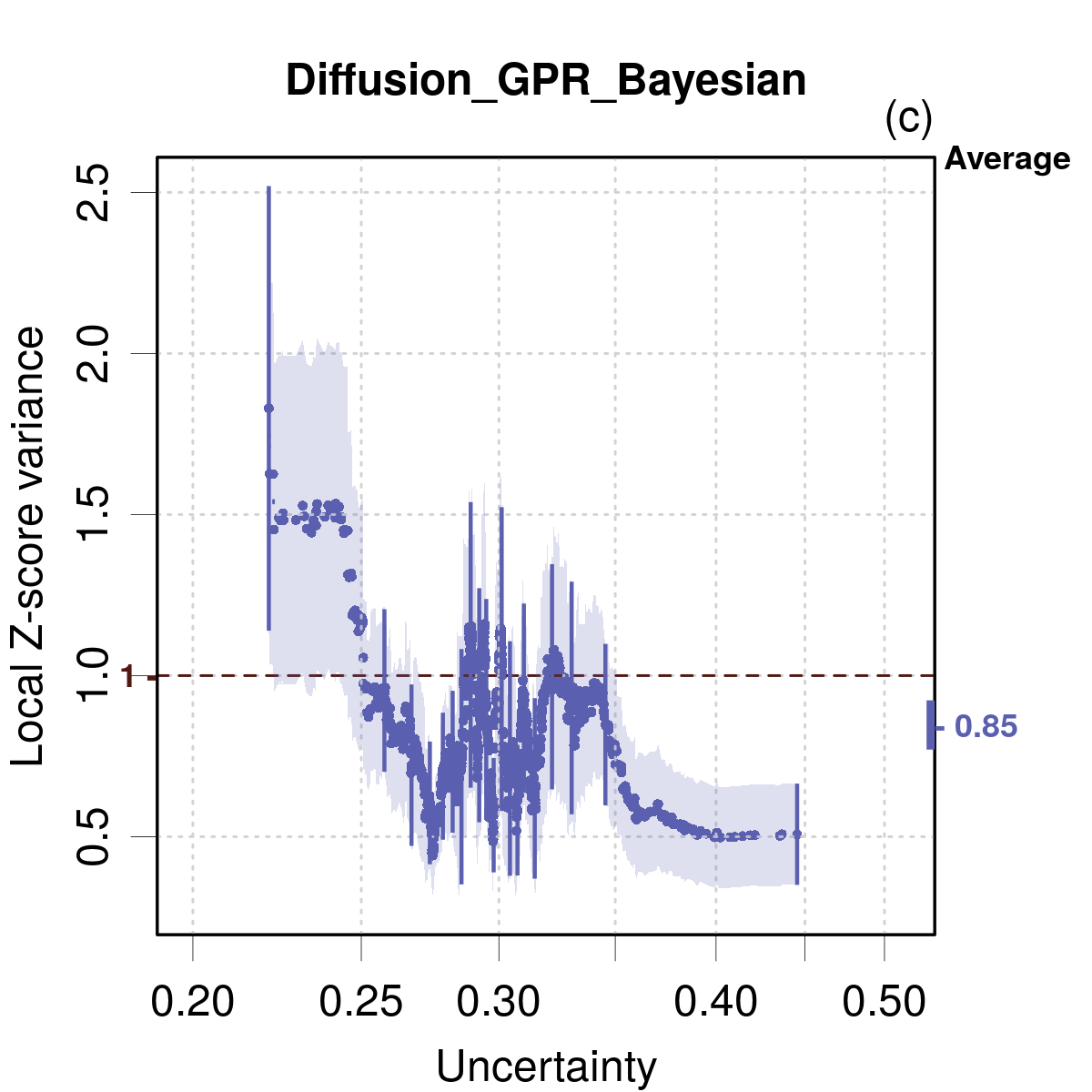}\includegraphics[height=8cm]{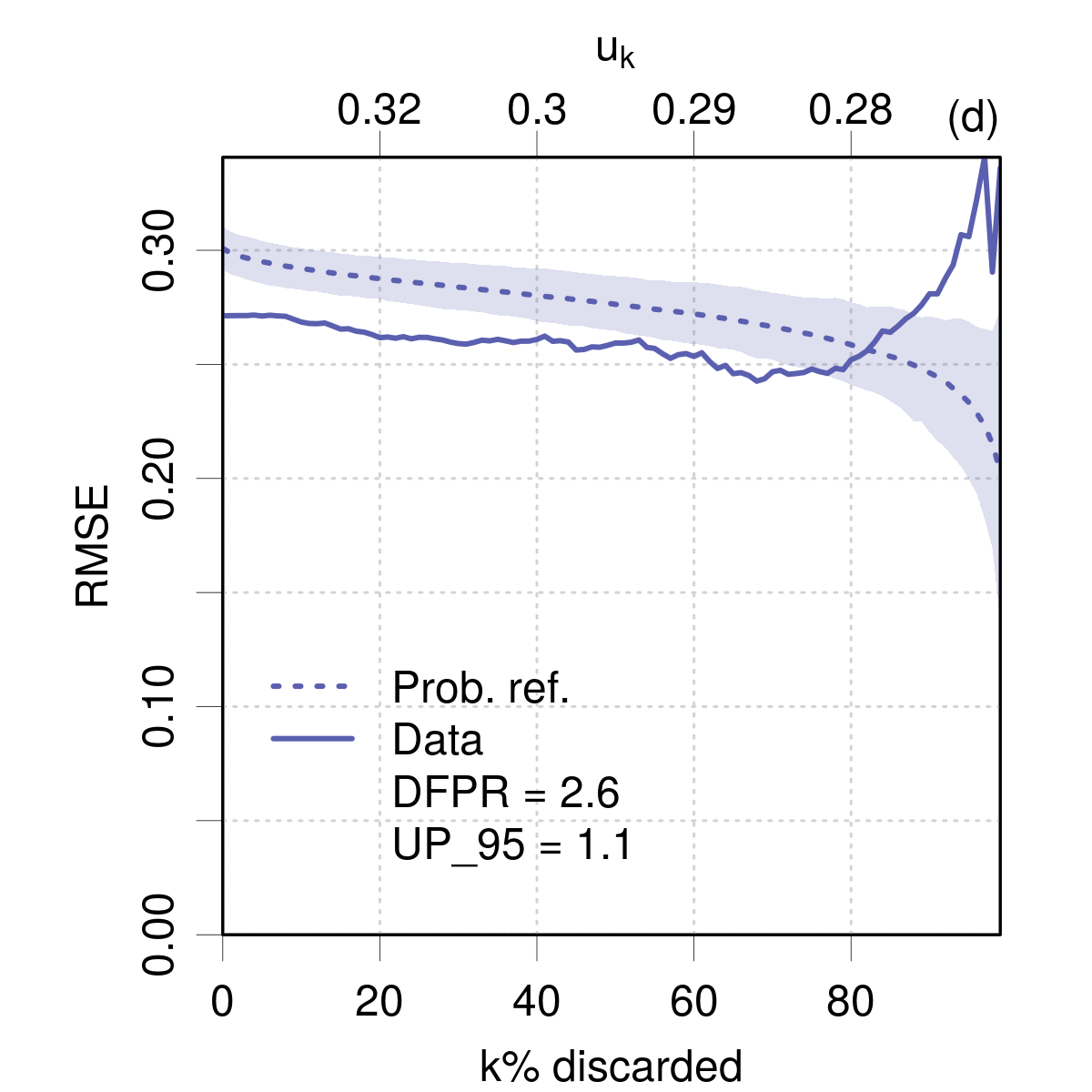}
\par\end{centering}
\caption{\label{fig:PAL1}Validation of calibrated uncertainties for the LR
and GPR\_Bayesian ML methods applied to the Diffusion dataset of Palmer
\emph{et al.}\citep{Palmer2022}. }
\end{figure}

For the Diffusion\_LR (Line Ridge regression) set {[}Fig.\,\ref{fig:PAL1}(a,b){]},
calibration is not perfect (the confidence bar around the average
$\mathrm{Var}(Z)$ value does not overlap the unity), and one observes
notable deviations of the LZV values for the larger uncertainties
(above $u_{E}=0.5$), albeit these deviations do no exceed a factor
two. This is translated by a confidence curve lying above the probabilistic
reference in the first 20-30 percent, after which the curve falls
back onto the reference. Here again, one cannot conclude to a perfect
calibration. The DFPR statistic leads to reject tightness.

The Diffusion\_GPR\_Bayesian (Gaussian PRocess) case {[}Fig.\,\ref{fig:PAL1}(c,d){]}
presents a somewhat pathological case, where the uncertainties cover
a small range. In addition to a sub-optimal calibration ($\mathrm{Var}(Z)=0.89)$,
the LZV analysis reveals tightness problems at both extremities of
the $u_{E}$ range. The confidence curve confirms the residual overestimation
of uncertainties (it lies below the reference until $k\simeq80\thinspace\%$).
Moreover, the initial 10\,\% of the confidence curve are flat and
it starts an erratic non-decreasing trajectory above $k\simeq65\thinspace\%$.
Clearly, these uncertainties are not reliable and should not be trusted
for active learning. 

\begin{figure}[t]
\noindent \begin{centering}
\includegraphics[height=5.5cm]{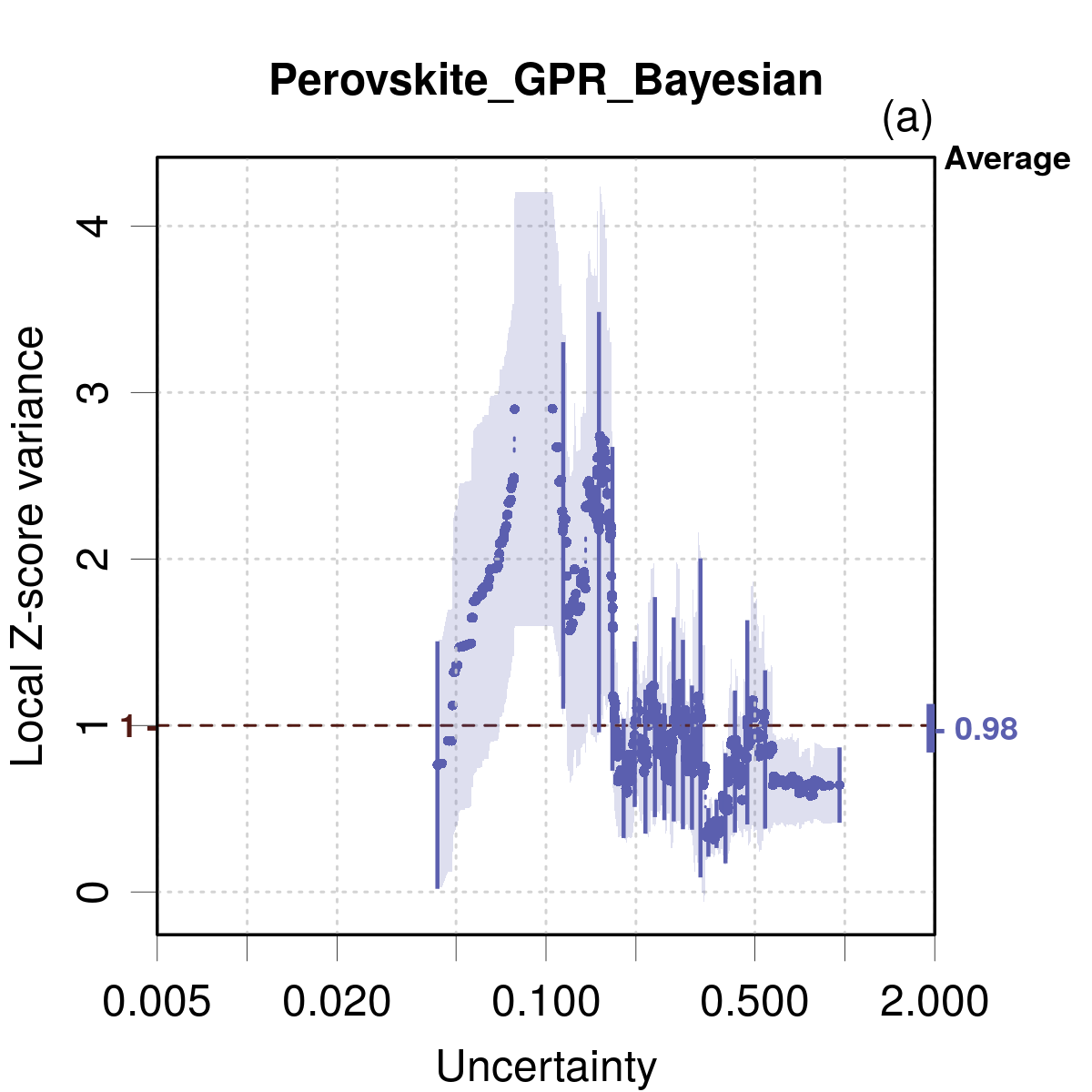}\includegraphics[height=5.5cm]{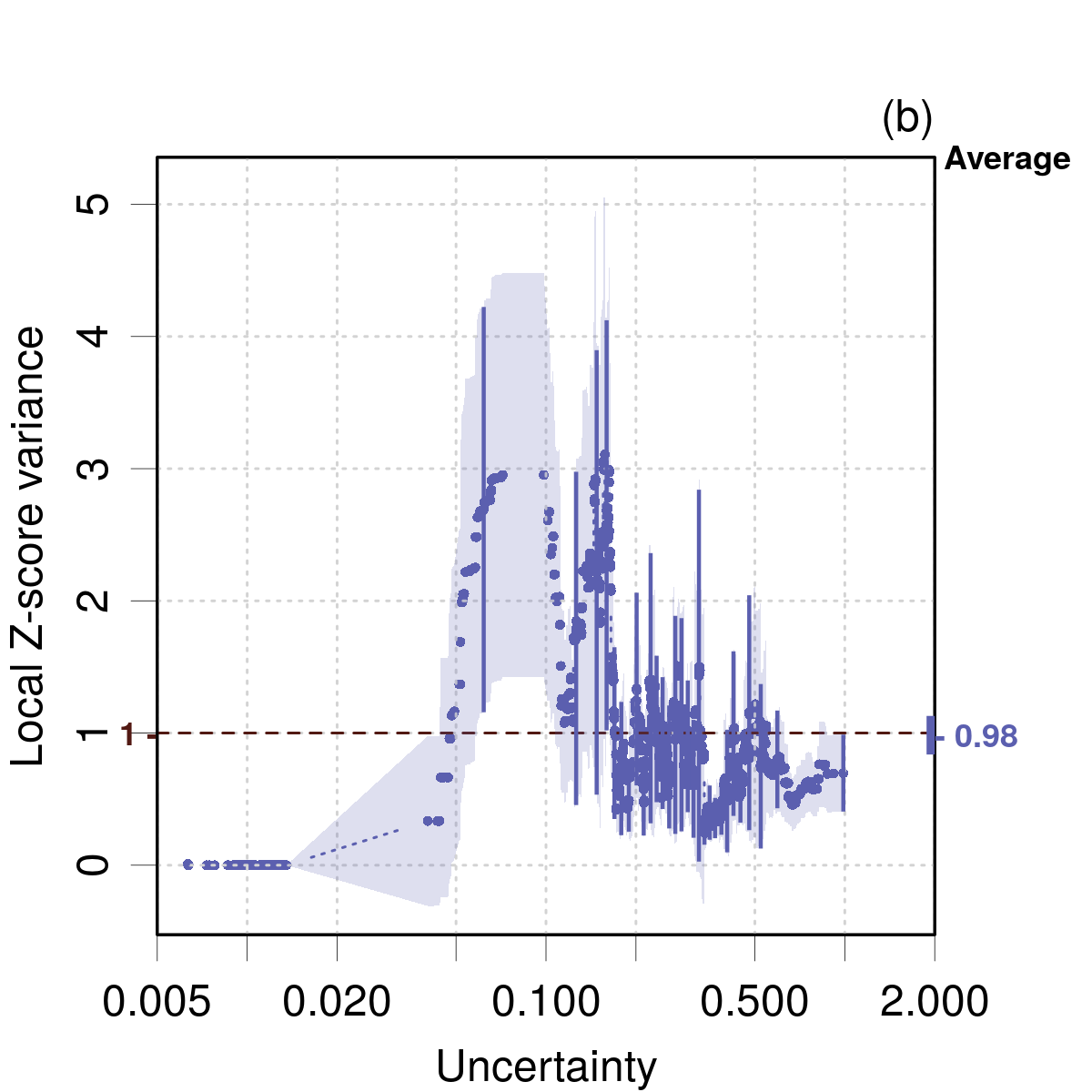}\includegraphics[height=5.5cm]{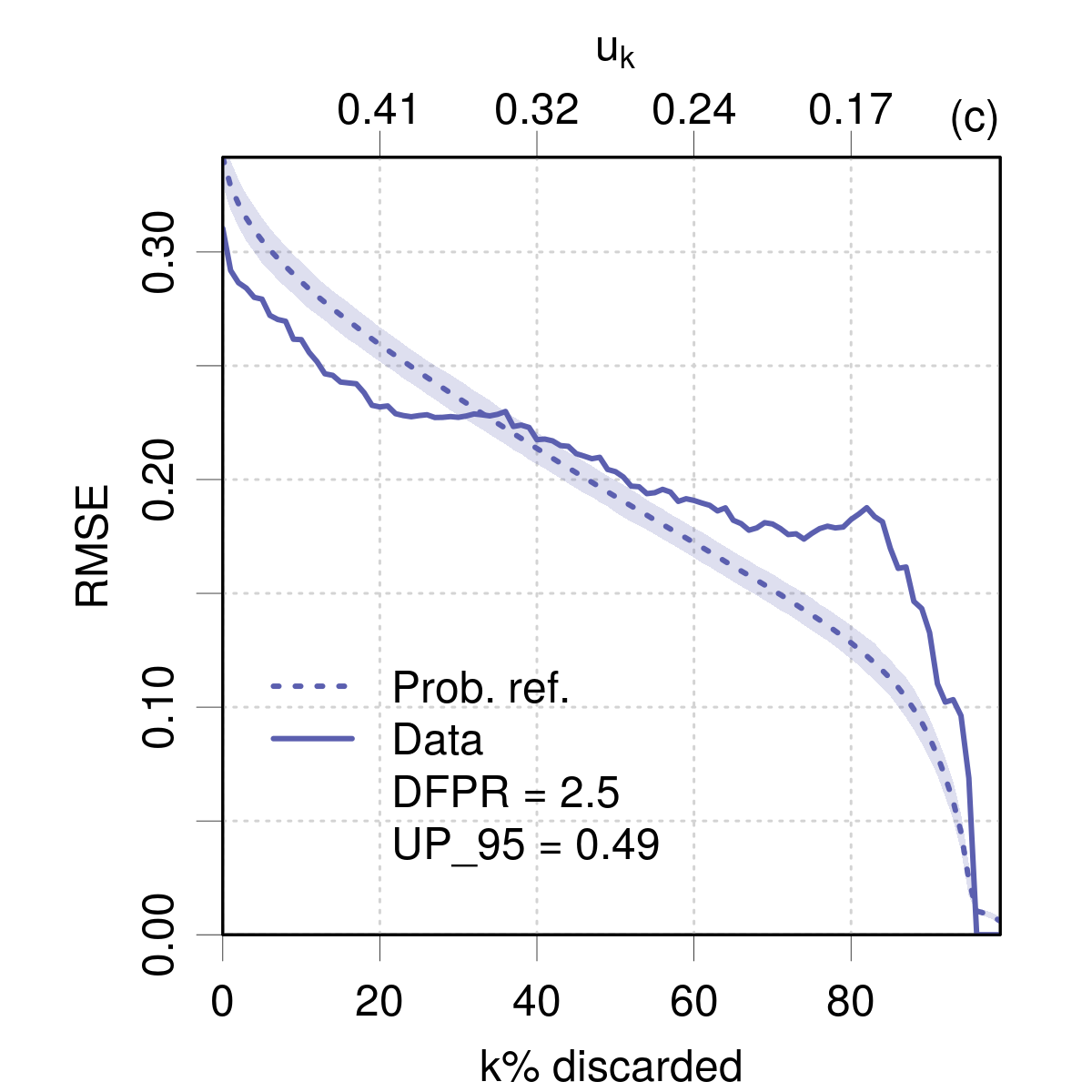}
\par\end{centering}
\caption{\label{fig:PAL2}Validation of calibrated uncertainties for Perovskite\_GPR\_Bayesian
dataset of Palmer \emph{et al.}\citep{Palmer2022}. }
\end{figure}

Let us now consider the Perovskite\_GPR\_Bayesian case {[}Fig.\,\ref{fig:PAL2}(a-c){]}
for which the confidence curve {[}Fig.\,\ref{fig:PAL2}(c){]} presents
a sharp drop to 0 before $k=100\thinspace\%$. In contrast, the smaller
uncertainties appear to be well calibrated in a standard LZV analysis
with 20 bins {[}Fig.\,\ref{fig:PAL2}(a){]}. Using a higher resolution
for the LZV analysis (30 bins) reveals the problem: a $z$-scores
variance of nearly 0 is observed for uncertainties around $u_{E}=0.01$.
Inspection of the dataset shows that these uncertainties are associated
with a set of errors of tiny amplitude ($|E|\le10^{-8}$). This ``nugget''
does not appear in the original analysis by Palmer \emph{et al.}\citep{Palmer2022}
who checked calibration by a reliability diagram with 15 bins. It
probably deserves further inquiry, as one might wonder about its leverage
in the calibration process. Nevertheless, both the LZV analysis and
confidence curves conclude to an absence of tightness, despite the
correct average calibration.

It is notable here that, as it does not depend on a binning strategy,
the confidence curve is more likely to reveal abnormal features of
the dataset that would be averaged out in a LZV analysis (or a reliability
diagram) with insufficient resolution. 

\section{Discussion\label{sec:Discussion}}

The most direct diagnostic feature of a confidence curve is that it
has to be continuously decreasing to reveal a proper association between
the errors and a UQ metric. This enables to use a UQ metric as a proxy
to detect the risk of large errors. To be able to use confidence curves
for UQ datasets comparisons or for calibration inquiries, one needs
a reference curve. I discuss below the main points relevant to the
choice of a reference curve, the design of a meaningful confidence
curve analysis, and the compared advantages of confidence curves versus
other calibration validation methods. 

\paragraph{The oracle is irrelevant to assess the calibration of variance-based
UQ metrics.\label{subsec:The-oracle-is}}

I have shown above that the often used oracle is not relevant to evaluate
confidence curves for variance-based UQ metrics (typically the uncertainty
as the standard deviation of an uncertain variable) for two main reasons:
(1) it corresponds to an unsuitable distribution of errors; and (2)
it does not depend on the scale of the UQ metric. Therefore, I strongly
recommend against its use in the variance-based UQ framework. Similarly,
it should not be used to compare the performances of amplitude-based
and variance-based UQ metrics.

\paragraph{The probabilistic reference enables to assess tightness (and sometimes
calibration) of uncertainties.}

By design, the probabilistic reference curve enables to test if $|E|$
and $u_{E}$ are in a correct probabilistic association. If a non-normalized
confidence curve lies within the probabilistic reference band, one
gets similar confirmation of \emph{calibration }and\emph{ tightness}
to what would be obtained from a reliability diagram\citep{Levi2020}
or LZV analysis.\citep{Pernot2022b} For instance, a reference curve
lying above the data confidence curve means that the uncertainties
are over-estimated, one can safely reject the calibration of the dataset.
Moreover, notable excursions of the confidence curve out of the probabilistic
reference band indicate a lack of tightness.\textcolor{orange}{{} }It
is not always possible to conclude about calibration. For instances,
in cases where the confidence curve lies around the probabilistic
reference but wanders out of the confidence band, one gets no direct
information about calibration. In such cases, a complementary calibration
test, such as $\mathrm{Var}(Z)$ is necessary.

\paragraph{When using a probabilistic reference, the best choice of statistic
is the RMSE.}

In order to validate calibration and tightness, the best configuration
is a non-normalized RMSE-based confidence curve. For the reference,
a normal generative distribution is a good default choice, unless
additional information is available. The reference curves based on
the MAE statistic are dependent on the chosen generative distribution,
which makes them less suitable, unless when using normalized confidence
curves.

\paragraph{The normalized confidence curves are less informative than the non-normalized
ones.}

The non-normalized statistic $c_{S}$ is easier to interpret than
$\tilde{c}_{S}$ and provides a richer diagnostic. The normalized
version is invariant by any monotonous transformation of the uncertainties
and scaling of the errors. Therefore, it cannot provide information
about calibration. If one uses a normalized CC, it is always necessary
to test \emph{average calibration}, for instance\emph{ }by checking
that $\mathrm{Var}(Z)\simeq1$ .\citep{Pernot2022b} Conditional to
a positive calibration test, the normalized confidence curve can help
to appreciate the tightness of uncertainties.

\paragraph{Comparison to other validation methods.}

Except for clear cut cases where a confidence curve does not cross
the probabilistic reference band, it does not provide unambiguous
information about average calibration, nor does it provide a quantitative
estimation. The proposed DFPR statistic does not escape to this ambiguity
and is interesting to confirm tightness. If tightness is rejected,
it cannot inform us about calibration. In this respect, the $\mathrm{Var}(Z)$
statistic is a useful/necessary complement. The LZV analysis provides
similar diagnostics about tightness as the confidence curve, with
the added value of a quantitative estimation. Moreover, the LZV analysis
can be performed against any quantity besides $u_{E}$, such as the
predicted values, which might provide better indices to locate and
solve tightness defaults. On the other hand, the applications above
show that an advantage of the confidence curve over the LZV and reliability
diagrams is that it is independent of a binning strategy. My recommendation
would thus be to use in priority the confidence curve analysis including
a DFPR statistic, and if the diagnostic is not clear cut (either a
clear absence of calibration or a positive tightness diagnostic),
to use a LZV analysis for more quantitative information.

\section{Conclusion\label{sec:Conclusion}}

This study has shown that the oracle reference is unsuitable to evaluate
confidence curves for variance-based UQ metrics, and should be confined
to the analysis of amplitude-based ones. I introduced a probabilistic
reference which presents the added value to enable calibration/tightness
diagnostic, going much beyond the capacities of the purely ranking-based
oracle. A good setup for a CC analysis has been shown to be: a normal
generative distribution $D$ to estimate the samples necessary to
the reference, and the RMSE statistic to estimate non-normalized confidence
curves of the data and reference. Even if it has sometimes to be complemented
by more quantitative methods, the CC analysis is a very interesting
method to address UQ calibration/tightness validation.

\section*{Code and data availability\label{sec:Code-and-data}}

The code and data to reproduce the results of this article are available
at \url{https://github.com/ppernot/2022_Confidence/releases/tag/v1.0}
and at Zenodo (\url{https://doi.org/10.5281/zenodo.6793828}). The
\texttt{R},\citep{RTeam2019} \href{https://github.com/ppernot/ErrViewLib}{ErrViewLib}
package implements the plotCC and plotLZV functions used in the present
study, under version \texttt{ErrViewLib-v1.6} (\url{https://github.com/ppernot/ErrViewLib/releases/tag/v1.6}),
also available at Zenodo (\url{https://doi.org/10.5281/zenodo.7445577}).\textcolor{orange}{{}
}The \texttt{UncVal} graphical interface to explore the main UQ validation
methods provided by \texttt{ErrViewLib} is also freely available (\url{https://github.com/ppernot/UncVal}).

\bibliographystyle{unsrturlPP}
\bibliography{NN}

\end{document}